\documentclass[aps,prl,preprint,showpacs,amsmath,amssymb]{revtex4-1}
\usepackage{graphicx}
\usepackage{dcolumn}
\usepackage{bm}

\newcommand{\STO}{$\text{SrTiO}_3$}

\begin{document}

\title{Dynamic flexoelectric effect in perovskites from first principles calculations}

\author{Alexander Kvasov}
\affiliation{Ceramics Laboratory, Swiss Federal Institute of Technology (EPFL), CH-1015 Lausanne, Switzerland}

\author{Alexander K. Tagantsev}
\affiliation{Ceramics Laboratory, Swiss Federal Institute of Technology (EPFL), CH-1015 Lausanne, Switzerland}
\affiliation{Ioffe Physical-Technical Institute, 26 Politekhnicheskaya, 194021, St. Petersburg, Russia}

\date{\today}
\pacs{77.65.-j,77.65.Ly}

\begin{abstract}
Using the dynamical matrix of a crystal obtained from ab initio  calculations, we have evaluated for the first time the strength of the dynamic flexoelectric effect and found it comparable to that of the static bulk flexoelectric effect, in agreement with earlier order-of-magnitude estimates.
We also proposed a method of evaluation of these effects directly from the simulated phonon spectra.
This method can also be applied to the analysis of the experimental phonon spectra, being currently the only one enabling experimental characterization of the static bulk flexoelectric effect.
\end{abstract}

\maketitle

The flexoelectric effect, which is a polarization response to a strain gradient, has been attracting appreciable attention of theorists and experimentalists.
Being a higher-order effect with respect to piezoelectricity, it becomes appreciable at the nanoscale, where large strain gradients are expected.
Recent experimental studies attest to many flexoelectricity-driven phenomena.
For example, a strain gradient can work as an electric field via flexoelectric coupling: it can induce poling, switching, and rotation
of polarization \cite{Gruverman_2003,Lyahovitskaya_2005,Lee_2011,Lu_2012}.
It can create a voltage offset of hysteresis loops and smear the dielectric anomaly at ferroelectric phase transitions \cite{Catalan_2004}.
The flexoelectric effect actually represents 4 related phenomena: static and dynamic bulk flexoelectric
effects, surface flexoelectric effect and surface piezoelectricity \cite{Yudin_2013}.
Among the four effects only the static  bulk flexoelectric effect can be viewed  as a high-order analog of piezoelectricity.
It is the only effect that has been quantitatively assessed by theorists \cite{Maranganti_2009,Hong_2010,Ponomareva_2012,Hong_2011,Stengel_2013}.
In contrast, dynamic flexoelectricity has never been  quantitatively assessed either theoretically or experimentally.
Though, a  general theory of the effect has being offered \cite{Tagantsev_1986}.
Concerning the size of the effect, it was only demonstrated that the dynamic effect is expected to be comparable to the static bulk flexoelectricity \cite{Yudin_2013}.
Thus, one  raises the reasonable question of the real size of dynamic flexoelectricity.
It is an important question since nowadays papers on electromechanical modelling involving flexoelectricity always neglect the dynamic flexoelectric effect.
Another issue missing from the current state-of-the-art of flexoelectricity is a method of experimental evaluation of this effect.

In this Letter we present the results of quantitative modeling of the dynamic flexoelectric effect in cubic \STO{} (STO), the classical perovskite material with the most studied flexoelectric properties.
We will demonstrate that the size of the dynamic flexoelectric effect is indeed comparable to that of the static bulk flexoelectricity, identifying a situation where the dynamic effect is a few times larger than the static.
We will also offer a method for the evaluation of the dynamic flexoelectric effect from simulated phonon spectra; the method that can also be used for assessment of this effect using experimental phonon data.

In terms of electromechanical constitutive equations \cite{Yudin_2013}
\begin{equation}\label{eq_Ei}
E_i = \chi_{ij}^{-1} P_j - f_{klij} \frac{\partial u_{kl}}{\partial x_j} + M_{ij} \ddot{U}_j -
g_{ijkl} \frac{\partial^2 P_k}{\partial x_j \partial x_l} + \gamma_{ij} \ddot{P}_j,
\end{equation}
\begin{equation}\label{eq_elastic}
\rho \ddot{U}_i = c_{ijkl} \frac{\partial u_{kl}}{\partial x_j} + f_{ijkl} \frac{\partial^2 P_k}{\partial x_l \partial x_l} - M_{ji} \ddot{P}_j,
\end{equation}
the static and dynamic bulk flexoelectric effects are introduced via the \emph{flexocoupling tensor} $f_{klij}$ and tensor $M_{ij}$, which will be termed as \emph{flexodymanic tensor}, respectively.
Equations \eqref{eq_Ei} and \eqref{eq_elastic} are written in the Cartesian reference frame $x_i$, summation over repeating suffixes is accepted.
We use the following notation: $P_j$  - polarization, $E_j$  - macroscopic electric field, $U_j$  - elastic displacements, $u_{ij}=\frac{1}{2}(\partial U_i/\partial x_j+\partial U_j/\partial x_i)$  - strain tensor in the approximation of small strains.

When one is interested in the macroscopic (static or dynamic) response, one can neglect the last two terms in Eqs. \eqref{eq_Ei} and \eqref{eq_elastic}, which correspond to higher dispersions.
Then, for the case of the static strain gradient ($\ddot{U}_j=0$), defining the flexoelectric response under the condition of a vanishing macroscopic field \cite{Yudin_2013}, Eq. \eqref{eq_Ei} yields
\begin{equation}\label{static}
P_k=\mu_{ijkl}\frac{\partial u_{ij}}{\partial x_l}
\end{equation}
where the flexoelectric tensor reads as
\begin{equation}\label{mu_f}
\mu_{ijkl} = \chi_{kx} f_{ijxl}.
\end{equation}
Equations \eqref{static} and \eqref{mu_f} describe the static bulk flexoelectric effect.

In the dynamic case, eliminating $\ddot{U}_j$ between \eqref{eq_Ei} and \eqref{eq_elastic} one finds an alternative expression for the flexoelectric tensor
\begin{equation}\label{mu_tot}
\mu_{ijkl} = \chi_{kx} f_{ijxl}^{\textrm{tot}}.
\end{equation}
\begin{equation}\label{f_tot}
f_{ijxl}^{\textrm{tot}}= f_{ijxl}-\rho^{-1}M_{xm}c_{mlij}.
\end{equation}
Equation \eqref{f_tot} introduces the \emph{total flexocoupling tensor} $f_{ijxl}^{\textrm{tot}}$, where the term containing $\rho^{-1}M_{xm}c_{mlij}$ controls the contribution of the dynamic flexoelectricity to the total bulk flexoelectric response.
As is clear from Eq. \eqref{eq_Ei} the origin of the dynamic effect is an accelerated motion of the medium.
Since in an elastic wave $\ddot{U_j} \propto \frac{\partial u_{kl}}{\partial x_j}$, the corresponding polarization response is proportional to the strain gradient.

In this work we are interested in the evaluation of the dynamic bulk flexoelectric effect and its comparison with the static effect for an STO crystal, in the Born-charge approximation.
To do this, the tensors $M_{xm}$ and $f_{ijxl}$ were obtained using the expansion of the dynamical matrix $A_{i p,i' p'}(\vec{q})$ calculated with the long-range contribution of the macroscopic electric field being excluded \cite{Tagantsev_1986}.
Here, the suffixes $i$ and $i'$ enumerate the Cartesian components of the atomic displacements while $p$ and $p'$ - atoms in an elementary unit cell of the crystal; $\vec{q}$  denotes the wave-vector.
Specifically, one needs the coefficients of the following long-wavelength expansion of the dynamical matrix
\begin{equation}\label{eq_A_exp}
A_{i p,i' p'}(\vec{q}) = A_{i p,i' p'}^{(0)}+ A_{i p,i' p'}^{(1)j} q_j + \frac{1}{2} A_{i p,i' p'}^{(2)jl} q_j q_l+...,
\end{equation}
and the matrix $\Gamma_{i p, i' p'}$, that is the inverse of the singular matrix $A_{i p,i' p'}^{(0)}$ defined in a special way \cite{Born_1962}.
The results of Tagantsev's approach \cite{Tagantsev_1986}, written for  a cubic crystal of perovskite structure in the Born charge approximation,  can be summarised as follows
\begin{equation}\label{eq_M_micro}
M_{ij} = \frac{\chi^{-1}}{v} Q_{ix,p} \Gamma_{x p, j p'}( m_{p'} - \frac{1}{s} \sum_{p''} m_{p''}),
\end{equation}
\begin{equation}\label{eq_f_micro}
f_{ijkl} = \frac{\chi^{-1}}{v} Q_{kx,p} (N_{x,p}^{ijl} + N_{x,p}^{jil} - N_{x,p}^{lij}),
\end{equation}
\begin{equation}\label{eq_N}
N_{i,p}^{jkl} = \frac{1}{2}\sum_{p''} \Gamma_{i p, i' p'}(A_{i' p',j p''}^{(2)kl}-
\frac{\delta_{p' p''}}{s} \sum_{q' q''} A_{i' q',j q''}^{(2)kl}).
\end{equation}
where all summations are done over all $s$ atoms of the unite cell.
Here $v$ is the volume of the unit cell; $Q_{ij,p}$ and $m_p$ are the matrix of the Born charges and the mass  of $p$-th atom in the unit cell; $\chi$ is the only non-zero component (in a cubic crystal) of the tensor of dielectric susceptibility\cite{Born_1962}
\begin{equation}\label{chi}
\chi_{ij} = \frac{1}{v} Q_{ix,p} \Gamma_{x p, y p'}Q_{jy,p'}.
\end{equation}
Thus to find  the tensors $\chi_{xm}$,  $M_{xm}$ and $f_{ijxl}$, only the matrices $A_{i p,i' p'}^{(0)}$ and  $A_{i p,i' p'}^{(2)jl}$ (available once the dynamical matrix $A_{i p,i' p'}(\vec{q})$  is calculated), the matrix of the Born charges $Q_{ix,p}$ and the volume of the unit cell are needed.

We found $A_{i p,i' p'}(\vec{q})$ and $Q_{ix,p}$ using first principles calculations.
Since, within zero Kelvin DFT calculations, the cubic structure of STO is unstable (negative squares of phonon frequencies for the soft-mode phonons) we performed our calculations under a hydrostatic pressure of 78 kBar to stabilize it, corresponding to approximately 1\% elastic strain.
Since the tensors $M_{xm}$ and $f_{ijxl}$ are not critically small, in view of the Landau theory arguments and atomic estimates, such strain is expected to induce some 1\% modifications of these tensors.
On the same lines, the result of zero-Kevin calculations should give a reasonable approximation of the finite-temperature values of these tensors.
Such an approach is justified by the experience of zero-Kevin calculations  of elastic constants.
This way, we believe that the estimates obtained here are valid for realistic finite temperature and pressure.

The evaluation of the contribution of dynamic flexoelectricity to the total flexocoupling tensor $f_{ijxl}^{\textrm{tot}}$, as is clear from Eq. \eqref{f_tot}, requires the values of the elastic constants $c_{mlij}$ which we also evaluated using the dynamical matrix, specifically for perovskite cubic structure using the relationship \cite{Born_1962}
\begin{equation}\label{eq_c_micro}
c_{ijkl} = \frac{1}{2v} \sum_{p,p'} A_{i p,k p'}^{(2)jl}.
\end{equation}

The technical details of the calculations are as follows.
The calculations were done for cubic STO exploiting the Quantum ESPRESSO (QE) \cite{QE_2009} ab initio package within the GGA PBE exchange-correlation functional with ultrasoft pseudopotentials \cite{pslib}.
We have used an automatically generated uniform 16x16x16 grid of k-points, the kinetic energy cutoff for wavefunctions was 80 Ry.
The calculations of phonon spectrum are done using Density Functional Perturbation Theory (DFPT) \cite{Giannozzi_DFPT} as implemented in the PHonon code.
The energy threshold for self-consistency was chosen to be equal to $10^{-20}$ Ry with the help of convergence tests for gamma point phonons.

To find the independent components of tensors $\chi_{ij}=\chi\delta _{ij}$, $M_{ij}=M\delta _{ij}$,  $f_{ijkl}$, and $c_{ijkl}$  ($\chi$, $M$, $f_{11}$, $f_{12}$, $f_{44}$, $c_{11}$, $c_{12}$, $c_{44}$), we evaluated the wave-vector dependence of the dynamical matrix $A_{i p,i' p'}(\vec{q})$ for the [100] and [110] directions of the reciprocal space.
This provided us with the components of the $A_{i p,i' p'}^{(0)}$ and  $A_{i p,i' p'}^{(2)jl}$ matrices needed to finalize the calculations using Eqs. \eqref{eq_M_micro}-\eqref{eq_c_micro}.

The results of the calculations are presented in  Tables \ref{table_DM} and \ref{table_ph}.
A remarkable observation, which  forms the main message of this Letter, is a very strong renormalization of the flexocoupling tensor by dynamic flexoelectricity: it is seen that the corresponding components of the tensors $f_{ijkl}$ and  $f_{ijkl}^{\textrm{tot}}$ can drastically differ.
This implies that the traditional neglect of dynamic flexoelectricity in the dynamic electromechanical simulations incorporating flexoelectricity is inadmissable.
Notably, the dynamic flexoelectricity can play an essential role in the frequency dependence of the electromechanical response close to the frequencies of mechanical resonances of the sample.
The point is that such response is expected to be sensitive to the dynamic flexoelectricity only above the relevant resonance frequency \cite{Yudin_2013}.

\begin{table}[!ht]
\centering
\begin{tabular}{|c|c|}
\hline
$M$ ($\times 10^{-8}\frac{\textrm{ Vs}^2}{\textrm{m}^{2}}$)  & $6 \pm 0.5$\\
\hline
$f_{11}$ (V) & $1.11 \pm 0.05$  \\
\hline
$f_{12}$ (V) & $-1.30 \pm 0.05$  \\
\hline
$f_{44}$ (V) & $-0.28 \pm 0.05$ \\
\hline
$f_{11}^\text{tot}$ (V) & $-2.1 \pm 0.1$  \\
\hline
$f_{12}^\text{tot}$ (V) & $-2.5 \pm 0.1$  \\
\hline
$f_{44}^\text{tot}$ (V) & $-1.5 \pm 0.1$ \\
\hline
$\chi/\epsilon_0$ & 2400 \\
\hline
\end{tabular}
\caption{Calculated static ($f$), dynamic ($M$), and total ($f^\text{tot}$) flexocoupling coefficients  for cubic \STO{} ($a=3.886$ \AA) under a pressure of 78 kBar using dynamical matrix.
$\chi/\epsilon_0$ is the relative dielectric susceptibility.}
\label{table_DM}
\end{table}

\begin{table}[!ht]
\centering
\begin{tabular}{|c|c|c|c|}
\hline
($10^{11} \frac{\textrm{N}}{\textrm{m}^2}$) & (a) Dyn. mat. & (b) Ph. disp. & (c) Exp. \\
\hline
$c_{11}$ & $2.77 \pm 0.05$ & $3.1 \pm 0.1$   & 3.16  \\
\hline
$c_{12}$ & $1.06 \pm 0.05$ & $0.9 \pm 0.1$ & 1.03  \\
\hline
$c_{44}$ & $0.92 \pm 0.05$ & $1.0 \pm 0.1$ & 1.22  \\
\hline
$|M| \frac{c_{44}}{\rho}$ (V) & $1.2 \pm 0.1$ & $1.1\pm0.4$ & \\
\hline
$|f_{44}^\text{tot}|$ (V) & $1.5 \pm 0.1$ & $1.5 \pm 0.2$ & 1.2-2.2 \\
\hline
$|f_{11}^\text{tot}-f_{12}^\text{tot}|$ (V) & $0.2 \pm 0.1$ & $< 0.5$ & 1.2-1.4 \\
\hline
\end{tabular}
\caption{Material parameters for cubic \STO{} ($a=3.886$ \AA) under a pressure of 78 kBar obtained using the dynamical matrix and from analysis of phonon dispersion curves.
$c$ is the stiffness tensor.
$f^\text{tot}$ is defined in Eq. \eqref{eq_f_eff_tot}.
$\rho$ is the density ($\rho=5174 \frac{\textrm{kg}}{\textrm{m}^3}$) calculated by the mass of atoms in the unit cell divided by the cell volume.
(a) Calculated from dynamical matrix.
(b) Obtained from the simulated phonon dispersion curves.
(c) Obtained from the experimental phonon dispersion curves \cite{Landolt-Bornstein,Vaks_1973,Tagantsev_2001}.}
\label{table_ph}
\end{table}

The results of our calculations of the dynamical matrix provide us with an alternative way of evaluating some components of the material tensors of STO directly from the phonon dispersion curves.
This will provide a cross-check  of the above results.
In addition, the method presented below will offer a way of evaluating these tensors from the experimental phonon spectra of the material.
We suggest fitting the long-wavelength part of the low-energy spectrum of the crystal to that obtained from the continuum theory, incorporating dynamic flexoelectricity.

The suggested approach uses the spectra of transverse acoustic (TA) and soft-mode transverse optic (TO) branches for the high symmetry [100], [110], and [111] directions of wavevector.
In this case, the phonons are not accompanied by any wave of the electric field and the continuum  theory equations,  Eqs. \eqref{eq_Ei} and \eqref{eq_elastic} with $E_i=0$, yield the following dispersion equation for the frequency $\omega$ of the transverse modes \cite{Yudin_2013}:
\begin{align}
& (\omega^2 - \omega_A^2)(\omega^2 - \omega_O^2) = \frac{(\omega^2 M - q^2 f_\text{eff})^2}{\rho \gamma}, \label{eq_omega2}  \\
& \omega_A^2 = \frac{c_\text{eff} q^2}{\rho}, \quad
  \omega_O^2 = \frac{\chi^{-1} + g_\text{eff} q^2}{\gamma} \label{eq_omega2-2}
\end{align}
where the coefficients $f_\text{eff}$, $c_\text{eff}$, and $g_\text{eff}$ are defined according to  the direction:
\begin{align}
& [100]: c_\text{eff}=c_{44}, \quad f_\text{eff}=f_{44}, \quad g_\text{eff}=g_{44} \label{100}\\
& [110]:
\begin{cases}
& c_\text{eff}=\frac{1}{2}(c_{11}-c_{12}) \\
& f_\text{eff}=\frac{1}{2}(f_{11}-f_{12}) \\
& g_\text{eff}=\frac{1}{2}(g_{11}-g_{12})\label{110}
\end{cases} \\
& [111]:
\begin{cases}
& c_\text{eff}=\frac{1}{3}(c_{11}-c_{12}+c_{44}) \\
& f_\text{eff}=\frac{1}{3}(f_{11}-f_{12}+f_{44}) \\
& g_\text{eff}=\frac{1}{3}(g_{11}-g_{12}+g_{44}) \label{111}
  \end{cases},
\end{align}
and $\gamma$ can be found from the frequency of the lowest TO mode at $\Gamma$-point and the susceptibility, namely \cite{Vaks_1973,Hehlen_1998,Farhi_2000}
\begin{equation}
\label{gamma}
\gamma = \frac{1}{\omega _O^2(0) \chi}.
\end{equation}

The roots of Eq. \eqref{eq_omega2} with coefficients defined by Eqs. \eqref{100} and \eqref{111} yield the frequencies of the corresponding double degenerate TA and TO modes.
The roots of Eq. \eqref{eq_omega2} with coefficients defined by Eqs. \eqref{110} yields the frequencies of the TA and TO modes that are polarized along the [$\overline{1}10$] direction.

One can perform the analysis of the dispersion of the acoustic branch by using Eq. \eqref{eq_omega2} and get information on the parameters $f_\text{eff}$ and $M$.
In the case of weak interaction between the branches (lowest order in $q$), the relative shift of the acoustic branch can be found as
\begin{equation}\label{eq_omega2_f_tot}
\omega^2 - \omega_A^2 = - \frac{\left(  q^2 f_\text{eff}^\text{tot} \right)^2}{\rho \gamma (\omega_O^2 - \omega_A^2)},
\end{equation}
where $f_\text{eff}^\text{tot}$ is defined as
\begin{equation}\label{eq_f_eff_tot}
f_\text{eff}^\text{tot} =  f_\text{eff}- M \frac{c_\text{eff}}{\rho},
\end{equation}
consistent with  Eq. \eqref{f_tot}.
As is clear from Eq. \eqref{eq_omega2_f_tot}, the deviation of the TA branch from the linear dispersion corresponds to the difference $\omega^2 - \omega_A^2$ proportional to $(f_\text{eff}^\text{tot}q^2)^2$.
The repulsive character of the interaction between the TA and TO branches (or lowering of the acoustic branch) is seen from the sign of rhs in expression \eqref{eq_omega2_f_tot}.
As for its magnitude, it is mainly controlled by the total flexocoupling coefficient \eqref{eq_f_eff_tot}, which has both dynamic and static contributions.
One should note that, attempting to evaluate the flexocoupling tensor, such fit was done in several papers \cite{Vaks_1973,Hehlen_1998,Farhi_2000,Tagantsev_2001}, however neglecting dynamic flexoelectricity, which is unacceptable in view of the results presented above.
Thus, the lowest order in the $q^2$ analysis of $|\omega^2 - \omega_A^2|$ can only provide information about the total flexocoupling coefficient, not distinguishing between the static and dynamic contributions.
The distinction between $f$ and $f^\text{tot}$ can be obtained by expanding the solution to Eq. \eqref{f_tot} up to the next power in $q$:
\begin{equation}\label{eq_omega2_fit}
\omega^2 - \omega_A^2 = - \frac{\left(  q^2 f_\text{eff}^\text{tot} -
q^4 M \frac{\chi{ \left( f_\text{eff}^\text{tot} \right) }^2}{\rho}  \right)^2}{\rho \gamma (\omega_O^2 - \omega_A^2)},
\end{equation}
A form of \eqref{eq_omega2_fit} suitable for analysis reads
\begin{equation}\label{eq_fit}
\Psi(q^2)\equiv\frac{\sqrt{\rho \gamma(\omega_A^2 - \omega^2) (\omega_O^2 - \omega_A^2)}}{q^2}
= \left| f_\text{eff}^\text{tot}-  q^2 M \frac{\chi{ \left( f_\text{eff}^\text{tot} \right)}^2}{\rho} \right|.
\end{equation}
It is clear that by fitting $\Psi(q^2)$ to a linear function of $q^2$ one can get information on  $M$ and $f_\text{eff}^\text{tot}$.
However, the resulting information on $M$ and the components of the flexocoupling tensor is limited.
Only the absolute values of $M$,  $f_{11}- f_{12}$, and  $f_{44}$ can be provided from such analysis.
We have checked that the analysis of the phonon spectrum for other directions than those addressed above unfortunately does not provide any additional information on the components of the flexocoupling tensor.
Actually, all the information can be obtained from the dispersion curves for any two of the [100], [110], and [111] wavevector directions, while the third direction may serve as a crosscheck and an evaluation of the inaccuracy of the method.

\begin{figure}[!ht]
\centering
\includegraphics[width=0.4\textwidth]{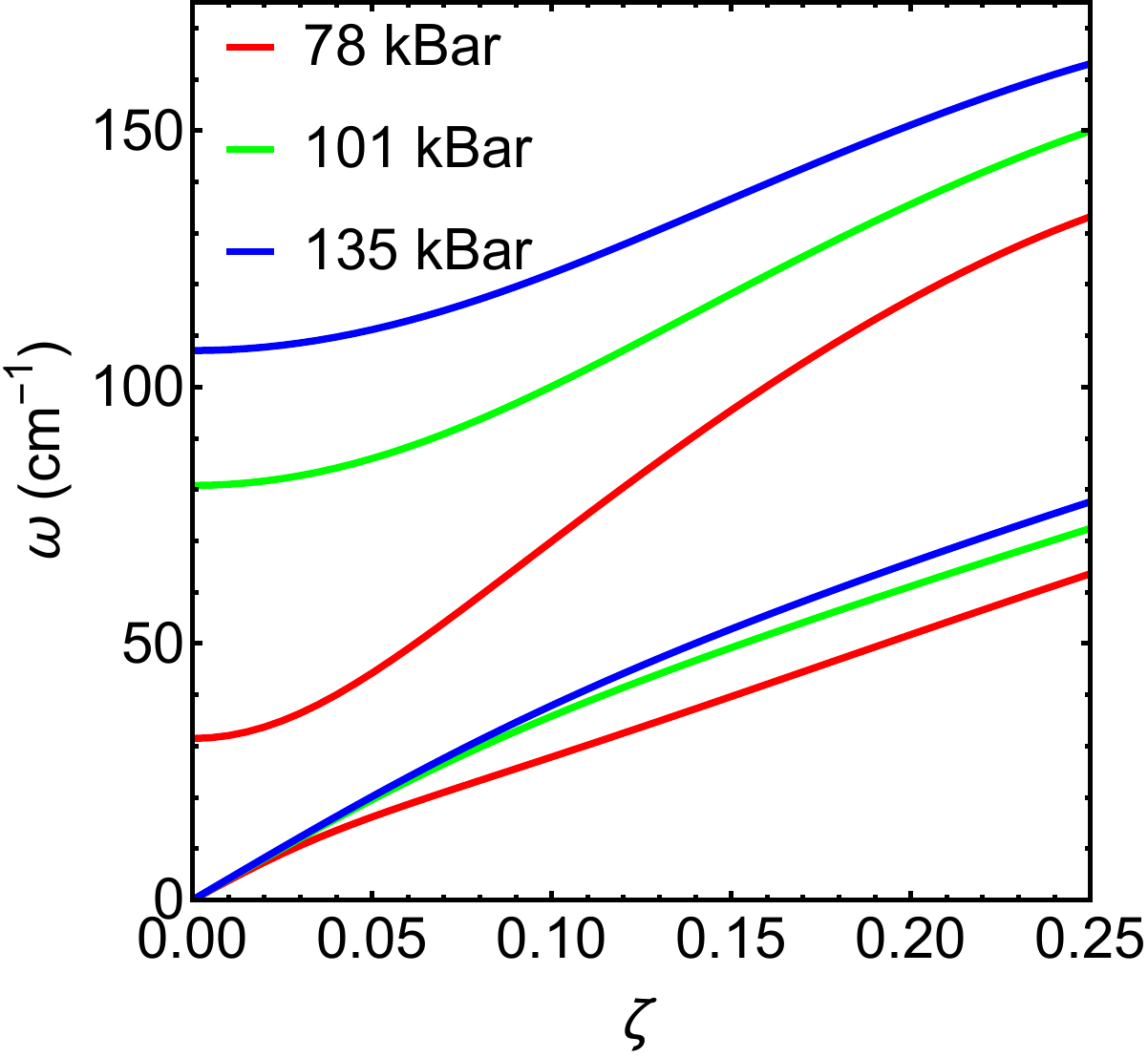}
\caption{Calculated phonon dispersion of transverse acoustic (TA) and soft-mode transverse optic (TO) branches with wavevector $q=\frac{\pi}{a} (\zeta,0,0)$ for cubic \STO{} stabilized by pressure.
The red lines correspond to lower pressure ($p=78$ kBar, $a=3.886$ \AA) when the flexoelectric interaction between the TO and TA branches is strong. Here one observes lowering of the TA caused by flexoelectric coupling.
The blue lines correspond to high pressure ($p=135$ kBar, $a=3.85$ \AA) when the interaction is weak in view of the large distance between the TA and TO branches.
One can also observe lowering of the TO branch and change of dielectric susceptibility: for low pressure (red line), we have $\omega _O(0) = 31 \textrm{cm}^{-1}$ and $\chi/\epsilon_0 = 2400$, whereas for high pressure (blue line), $\omega _O(0) = 107 \textrm{cm}^{-1}$, and $\chi/\epsilon_0 = 220$.}
\label{fig_ph_100_p}
\end{figure}

We have applied the described above method to evaluate the flexocoupling and flexodynamic tensors directly from  the simulated phonon spectra.
The phonon spectra were obtained for the [100], [110], and [111] wavevector directions on 10x10x10 automatic Monkhorst-Pack grid of q-points, in the framework already discussed above.
The calculations were performed under stabilizing pressures.
An example of the TA and soft mode TO spectra are shown in Fig. \ref{fig_ph_100_p}.
The TA-TO repulsion effect caused by the flexoelectric coupling is clearly seen in this figure.
To get information on the $M_{ij}$ and $f_{ijkl}$ tensors, the long-wavelength part of the spectra was calculated under a pressure of 78 kBar where the inter-mode coupling is a maximum.

\begin{figure}[!ht]
\centering
\includegraphics[width=0.4\textwidth]{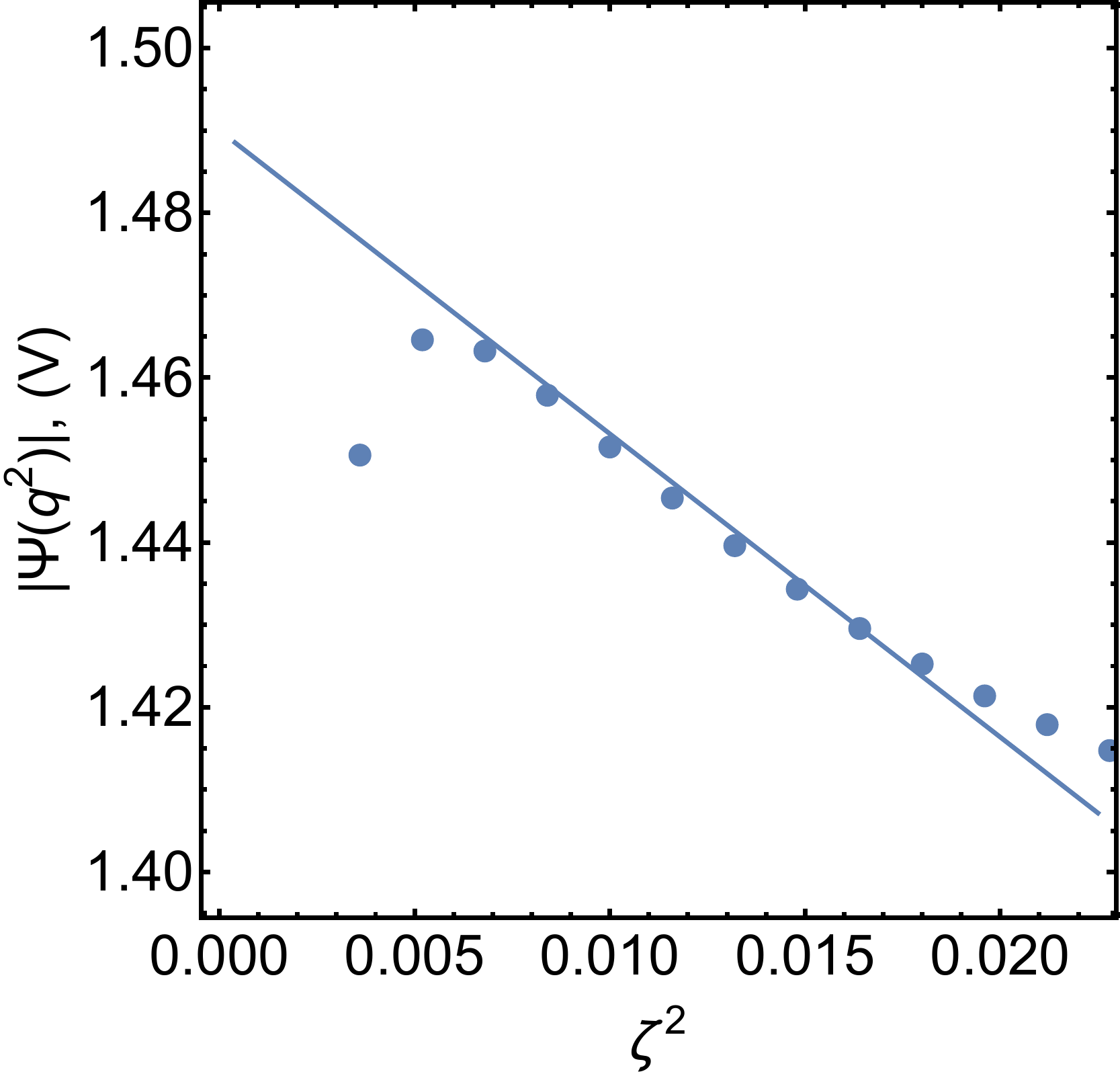}
\caption{Module of $\Psi (q^2)$ \eqref{eq_fit} (dots) and fitting line, which was used to obtain $f_\text{eff}^\text{tot}$ and $M$.
To evaluate $\Psi (q^2)$, the frequencies of the  transverse acoustic (TA) and soft-mode transverse optic (TO) branches entering in Eq. \eqref{eq_fit} were calculated for the [100] direction of wave vector, $q=\frac{2 \pi}{a} (\zeta,0,0)$, for cubic \STO{} stabilized by pressure.}
\label{fig_fit}
\end{figure}

As a first step, $M$ and $f_\text{eff}^\text{tot}$ were obtained from a linear fit for $\Psi(q^2)$ as a function of $q^2$ for the [100] and [110] directions.
An example of such a fit is shown in Fig. \ref{fig_fit}.
It is worth mentioning that the reliable evaluation of  the flexoelectric parameters directly from the spectrum is a challenging computational task.
The problem is that, on one hand, one is interested in the long-wavelength limit of the spectrum, on the other hand, close to the $\Gamma$-point the violation of the acoustic sum rules corrupts the simulated spectrum \cite{Giannozzi_DFPT}.
Thus, for this evaluation, only a limited interval of the eigenvectors should be used for the fit, as it was done in Fig.\ref{fig_fit}.
Here, we used the simulated TA and TO frequencies for $\omega^2$ and $\omega_O^2$, respectively, $\rho=5174 \frac{\textrm{kg}}{\textrm{m}^3}$.
$\omega_A^2$ and $\gamma$ were found using Eqs. \eqref{eq_omega2-2} and \eqref{gamma} with the susceptibility and elastic constants obtained from the results of the same first principles calculations.
Then, based on the results of the fit, using \eqref{eq_f_eff_tot} we evaluated the values of $|M|$, $|f_{11}- f_{12}|$, and $|f_{44}|$, which are given in Table \ref{table_ph}.
The spectrum for the [111] direction was found in a reasonable agreement with results obtained for those for the [100] and [110] directions, enabling us to evaluate the inaccuracy of our calculations given in this table.

Table \ref{table_ph} gives a comparison between the results obtained with the two methods and  with the available experimental data.
Note that for the moment, experimental information is only available on some components of the tensor $f_{ijkl}^{\textrm{tot}}$ is available.
It is seen from Table \ref{table_ph} that the theoretical results yielded by the two methods are in good agreement with each other.
The agreement between the theory and the data obtained from the Brillouin and neutron scattering experiments \cite{Landolt-Bornstein,Vaks_1973,Tagantsev_2001,Hehlen_1998} is reasonable.

The proposed method of evaluating the components of  flexodynamic and flexocoupling tensors from the simulated spectra can also be applied for the determination of these components from the experimental phonon dispersion curves of materials exhibiting a ferroelectric soft mode.
Thus, having information about phonon dispersion, dielectric susceptibility, and density of the material, one can exploit the above described method to find the flexodymanic tensor and certain combinations of flexocoupling coefficients experimentally.
It is of importance to mention that, formally, this method can be applied to the phonon spectrum of any material, however in the absence of low-energy TO mode, it will not give reliable information on the sought tensors.
The point is that in such a situation the self-dispersion of the TA mode (which is always present and related to other effects) is expected to be of the same order of magnitude as the dispersion caused by flexoelectricity.

It should be noted that for the moment the proposed method seems to be the only one enabling the evaluation of the strength of the static bulk flexoelectric effect (i.e. flexocoupling tensor) since static experiments in a finite sample (like those done by Zubko et al \cite{Zubko_2007}) yield the sum of the static bulk and surface contributions, which, in general, are comparable \cite{Yudin_2013}.

In conclusion, using ab initio DFPT calculations, for strontium titanate, we have evaluated the strength of the dynamic flexoelectric effect and compared it to that of the static bulk flexoelectric effect.
We have found these effects to be of a comparable magnitude, in agreement with earlier order-of-magnitude estimates \cite{Tagantsev_1986}.
We identified the situation where the dynamic effect is a few times stronger.
Taking into account that, currently, all experimental studies and simulations involving flexoelectricity ignore the dynamic effect, we believe that our findings are of importance.
We have also offered a method for extracting information on flexodynamic and flexocoupling tensors using a simulated phonon spectrum.
This method can be applied to the treatment of experimental phonon spectra, being currently the only method enabling an experimental evaluation of  the size of the bulk flexoelectric effect.

\begin{acknowledgments}
This project was supported by the grant of Swiss National Science Foundation, No. 200020-144463/1, and by the grant of the government of the Russian Federation, No. 2012-220-03-434.
\end{acknowledgments}

\end{document}